\documentstyle[12pt,epsfig]{article}
\newcommand{\scaption}[1]{\caption{\protect{\footnotesize  #1}}}

\newcommand{\av}[1]{\mbox{$ \langle #1 \rangle $}}
\newcommand{\lsim}{\mbox{$~{\stackrel{<}{\scriptstyle{ \sim}}}~$}}
\newcommand{\gsim}{\mbox{$~{\stackrel{>}{\scriptstyle{ \sim}}}~$}}
\newcommand{\qprime}{\mbox{$Q'~$}}
\newcommand{\qprimesq}{\mbox{$Q'^2~$}}
\newcommand{\xprime}{\mbox{$x'~$}}

\newcommand{\qprimex}{\mbox{$Q'$}}
\newcommand{\qprimesqx}{\mbox{$Q'^2$}}
\newcommand{\xprimex}{\mbox{$x'$}}

\newcommand{\einst}{\mbox{$\varepsilon_I$}}

\newcommand{\xb}{\mbox{$x~$}}  
\newcommand{\xbx}{\mbox{$x$}}  

\newcommand{\siglim}{\mbox{$\sigma_{\rm lim}~$}}
\newcommand{\sigdis}{\mbox{$\sigma_{\rm DIS}~$}}
\newcommand{\flim}{\mbox{$f_{\rm lim}$}}
\newcommand{\siginst}{\mbox{$\sigma_I~$}}

\newcommand{\Qsq}{\mbox{$Q^2~$}}
\newcommand{\Qsqx}{\mbox{$Q^2$}}
\newcommand{\et}{\mbox{$E_T~$}}
\newcommand{\etx}{\mbox{$E_T$}}

\newcommand{\pt}{\mbox{$p_T~$}}
\newcommand{\ptx}{\mbox{$p_T$}}

\newcommand{\nmax}{\mbox{$n_{\rm max}~$}}

\newcommand{\dif}{\mbox{\rm d}}

\newcommand{\GeV}{\mbox{\rm ~GeV~}}
\newcommand{\GeVx}{\rm GeV}

\newcommand{\GeVsq}{\mbox{${\rm ~GeV}^2~$}}
\newcommand{\GeVsqx}{\mbox{${\rm ~GeV}^2$}}

\newcommand{\pbx}{\mbox{${\rm ~pb}$}}
\newcommand{\pbinvx}{\mbox{${\rm ~pb^{-1}}$}}

\begin{document}
%
%
%
%
%
%
%
\begin{titlepage}

%
\noindent
{\tt DESY 97-151    \hfill    ISSN 0418-9833} \\
{\tt MPI-PhE/97-18} \\
{\tt hep-ex/9708008} \\
{\tt August 1997}                  \\


\begin{center}
\vspace*{2cm}


{\bf  BOUNDS ON  QCD INSTANTONS FROM HERA}\\
%
\vspace*{2.cm}
{\bf T. Carli\footnote{carli@desy.de} 
and M. Kuhlen\footnote{kuhlen@desy.de}} \\ 
\vspace*{1.cm}
Max-Planck-Institut f\"ur Physik \\
Werner-Heisenberg-Institut \\
F\"ohringer Ring 6 \\
D-80805 M\"unchen  \\
Germany
\bigskip
\bigskip
\\

\vspace*{2cm}

\end{center}

\begin{abstract}

\noindent 

Signals for processes induced by QCD instantons are 
searched for in HERA data on the hadronic final state
in deep-inelastic scattering.
The maximally allowed fraction of  
instanton induced events is found at 95\% confidence level
to be on the percent level in the kinematic domain
$10^{-4} \lsim \xb \lsim 10^{-2}$
and $ 5 \lsim \Qsq \lsim 100$\GeVsqx.
The most stringent limits are 
obtained from the multiplicity distributions.
\vspace{1cm}

\end{abstract}
{\tt 
}

\end{titlepage}

\newpage

\section{Introduction}

\begin{figure}[h]
\begin{tabular}{ll}
\mbox{
   \epsfig{file=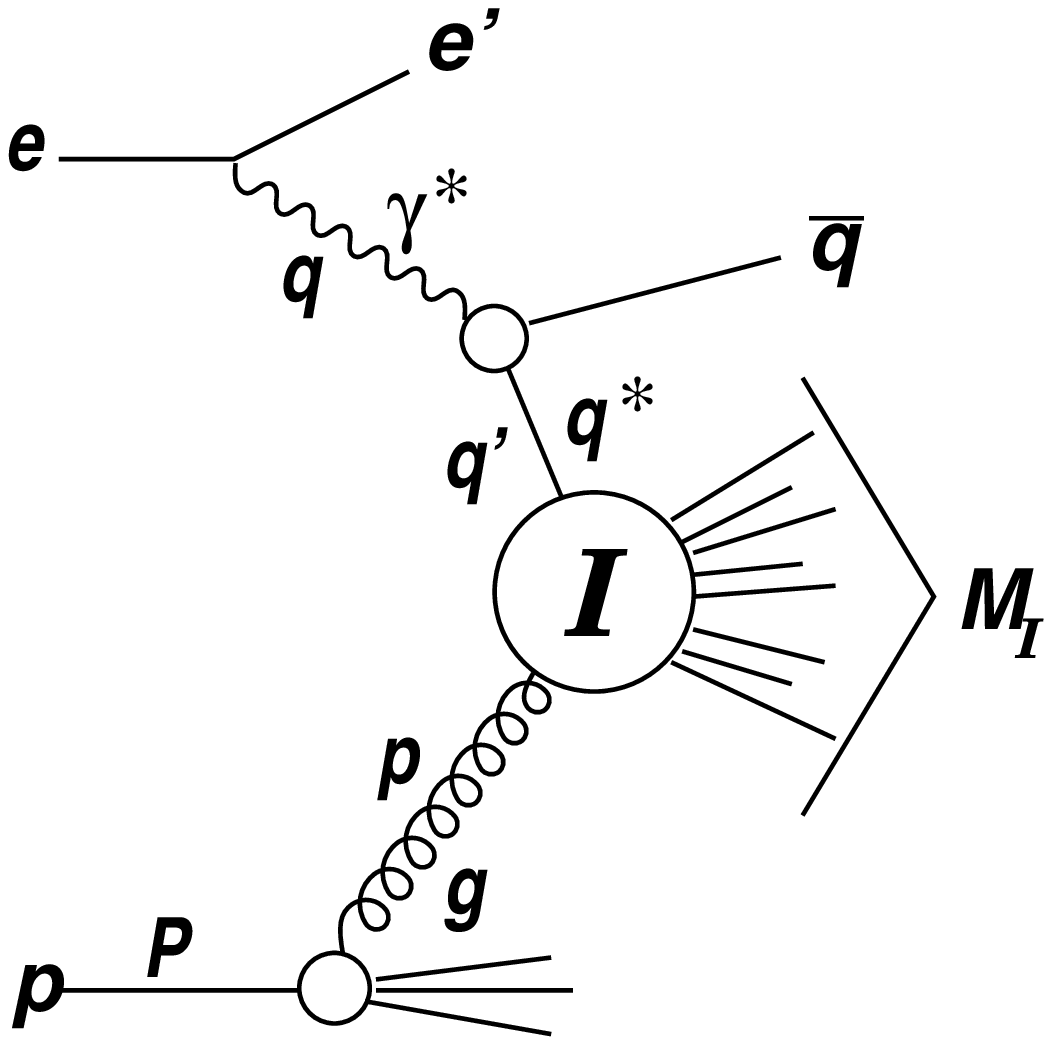,width=6.0cm,%
   bbllx=84pt,bblly=214pt,bburx=470pt,bbury=572,clip=}
}
&
\begin{tabular}{l}
 \vspace{-6.cm} \\
 DIS variables: \\
   $\Qsq= - q^2 $ \\
   $ x = \Qsq / (2 P \cdot q) $ \\
   $ W^2 = \Qsq (1 - x)/x$ \\ \\
 Variables of instanton subprocess: \\
 $\qprimesq=- q'^2 $ \\ 
 $x'= \qprimesq /(2 \; p \cdot q' ) $ \\
 $M_I^2= \qprimesq ( 1 - \xprime )/ \xprime $
\end{tabular}
\end{tabular} 
   \scaption{ Diagram of an instanton induced process in DIS,
    where a virtual photon with 4-momentum $q$ 
    emitted from the incoming electron fuses with
    a gluon with 4-momentum $p$ emitted from the proton
    with 4-momentum $P$. The virtual quark $q^*$ entering
    the instanton subprocess has 4-momentum $q'$. These
    4-momenta define the DIS variables \xbx, \Qsqx, and
    \xprime and \qprimesq characterizing the instanton subprocess.
    $W$ is the invariant mass of the hadronic final state, 
    and $M_I$ the invariant mass of the hadrons emerging from the 
    instanton subprocess.
   } 
   \label{fig:inst} 
\end{figure}

The
standard model contains processes which cannot be 
described by perturbation theory, and which violate
classical conservation laws like baryon and lepton number
in the case of the electroweak and chirality in the case of
the strong interaction \cite{inst:thooft}. 
Such anomalous processes are induced by instantons \cite{inst:belavin}.
At HERA, which collides $27.5$\GeV positrons on $820$\GeV protons,
QCD instantons may lead to observable effects
in the hadronic final state 
in deep-inelastic scattering (DIS) \cite{inst:balitsky1,
inst:vladimir,inst:paris95,inst:schremppdis96,
inst:heraws96,inst:moch,inst:yaroslavl,
inst:schremppdis97}.
Instantons isotropically decay into a high multiplicity state, 
consisting of gluons and all quark flavours which are kinematically 
allowed in each event (see Fig.~\ref{fig:inst}).
One expects therefore a densely populated
region in rapidity, other than the current jet, which is
homogeneously distributed in azimuth. 
The presence of strangeness and charm provides an additional signature.
From the analysis of $K^0$ yields \cite{h1:k0}
and multiplicity distributions \cite{h1:mult}
the H1 collaboration has put first limits on instanton production
for Bjorken~$x>0.001$.
They allow at most a few per cent admixture of instanton events
to normal DIS events.

We have investigated systematically the sensitivity of all available
HERA hadronic final state data to instanton production.
In this paper we analyze the data which are most sensitive
to instanton production, namely the
multiplicity distributions \cite{h1:mult},
the transverse energy flows \cite{h1:flow3},
and hard particle production \cite{h1:pt}.
We derive limits on instanton production,
improving existing limits by an order of magnitude, and
extend them into previously uncovered kinematic regions.

The paper is organized as follows: 
in chapter 2 the details of the 
Monte Carlo simulation of the hadronic final state are given,
in chapter 3 
the instanton phenomenology is described, 
in chapter 4 we discuss the HERA data in the light of
instanton production and present the new bounds, and finally
conclude in chapter~5.

\section{Monte Carlo simulation}

Predictions for the hadronic final state in normal and in instanton
induced DIS events are extracted from Monte Carlo generators, which
model the interaction.
They incorporate QCD evolution and parton radiation in
different approximations and utilize phenomenological models
for the non-perturbative hadronization phase.

\subsection{Standard QCD models}


The QCD Monte Carlo simulation program ARIADNE \cite{mc:ariadne}
uses the QCD matrix elements up to first order of the
strong coupling constant $\alpha_s$, with additional
multi-gluon emissions from a chain of independently radiating 
dipoles formed by the colour charges \cite{mc:dipole}.
The hadronization is performed with the LUND string model \cite{mc:string}
as implemented in JETSET \cite{mc:jetset}.
ARIADNE provides
an excellent description of all available HERA data on the hadronic
final state in DIS \cite{mc:heratune}. 

\subsection{Instanton Monte Carlo}

The Monte Carlo event generator QCDINS 
\cite{inst:qcdins,inst:ringwalddis97} is used to
model the hadronic final state of instanton induced processes
in DIS. 
It also predicts the total cross section in a restricted phase
space region.

Events induced by QCD instantons predominantly 
invoke a quark-gluon fusion process\footnote{Quark induced processes have
not yet been taken into account and are expected to be of minor importance.} 
as depicted in Fig.~\ref{fig:inst}.
The total cross section is given by a convolution of the
probability to find a gluon in the proton
$P_{g/p}$, the cross section 
$\sigma^{(I)}_{q^*g}(\xprimex,\qprimesqx)$ of the instanton induced
subprocess and the probability that a photon splits
into a quark-antiquark pair in the instanton background
$P^{(I)}_{q^*/\gamma^*}$ \cite{inst:yaroslavl,inst:moch}.
Besides the squared transverse momentum transfer \Qsq and
the Bjorken-\xb scaling variable, this scattering process is 
characterized by \qprimesqx, the virtuality of the quark ($q^*$),
and \xprimex, the Bjorken
scaling variable associated with the $q^*g$ subprocess
(see Fig.~\ref{fig:inst} for definition).

The cross section of the instanton induced subprocess is 
given by \cite{inst:yaroslavl}:
\begin{equation}
 \sigma_{q^*g}^{(I)}(\xprimex,\qprimesqx) \approx
 \frac{\Sigma(\xprimex)}{\qprimesqx}
{ \left(\frac{4 \pi}{\alpha_s(\mu(\qprimex))}\right)}^{\frac{21}{2}}
 {\rm exp} \left(\frac{- 4 \pi}{\alpha_s(\mu(\qprimex))} F(\xprimex) \right)
\label{eq:cross}
\end{equation}
where $\Sigma(\xprimex)$ and $F(\xprimex)$ are known functions 
of $\xprimex$ as long as \xprime is not too small, 
say $\xprime \gsim 0.2 $. $\mu(\qprimex)$ is the renormalisation
scale.
$F(\xprimex)$ modifies the exponential suppression factor
${\rm exp}(-4 \pi /\alpha_s)$ typical for tunneling processes 
and has its origin in multi-gluon emissions 
at high energies \cite{inst:ringwald}.
$F(\xprimex)$ is $1$ for $\xprimex = 1$ and is decreasing towards
small $\xprimex$. 
For $\xprimex \approx 0.2$, $F(\xprimex)$ is $\approx 0.5$. 
The assumed expression for $F(\xprimex)$ is considered to be a reasonable
estimate for $\xprimex \gsim 0.2$. The extrapolation to lower
values of \xprime is unreliable due to 
inherent ambiguities \cite{inst:yaroslavl}.
Moreover \qprimesq has to be large enough ($\qprimesq \gsim 25 \GeVsqx$) 
to allow (instanton) perturbation theory to be applied.
Another source of uncertainty comes from a residual renormalisation
scale dependence.

The total instanton induced $q^*g$ cross section 
decreases with increasing \qprimesq and
exponentially grows with decreasing \xprimex.
Since the center of mass energy of the $q^*g$ system 
is $M_I = \sqrt{\qprimesq (1-\xprimex)/\xprime}$, 
the expected event topology will be strongly influenced
by these two variables.
\xprime and \qprimesq are in principle observables measurable
from the hadronic final state. Here we consider instanton
events which are produced above various \xprime and \qprimesq
cut-offs.

From the instanton subprocess, $n_f$ quark-antiquark
pairs and $n_g$ gluons are being emitted isotropically
in the $q^*g$ center of mass system. 
$n_f$ is the number of flavours which are kinematically
allowed\footnote{In this analysis the maximally allowed number of
flavours was set to $5$. 
The mean multiplicity of charged
particles is reduced by $10$\% when only $4$ flavors
are considered.}.
Each event thus contains quarks of all kinematically allowed
flavors. 
The number of gluons $n_g$ in the instanton subprocess is generated
according to a Poisson distribution and is $\approx 2$ for 
$\xprimex \gsim 0.2$ and 
$\qprimesqx \gsim 25 \GeVsqx$\cite{inst:vladimir,inst:ringwalddis97}.

The flavour of the current quark 
(produced by the splitting of the photon
in the instanton background) is chosen at random.
Its $4$-vector is reconstructed from \xb and \Qsq using
a Sudakov decomposition.
The cross sections are calculated with massless quarks,
but the generated final state quarks are massive.

After assembling the hard instanton induced subprocess,
QCDINS allows for further gluon emission in the leading-logarithm 
approximation simulated with a coherent parton branching algorithm 
as implemented in HERWIG \cite{mc:herwig}. 
The transition from partons to the observable
hadrons is performed with the cluster fragmentation  model \cite{cluster},
where the primary hadrons are produced from an isotropic two
body decay of colour singlet parton clusters.
QED radiation is not implemented. The published data used in this
analysis had been corrected for QED radiation.


%
           
\section{The instanton induced final state}
\begin{figure}[h]
   \centering
   \epsfig{file=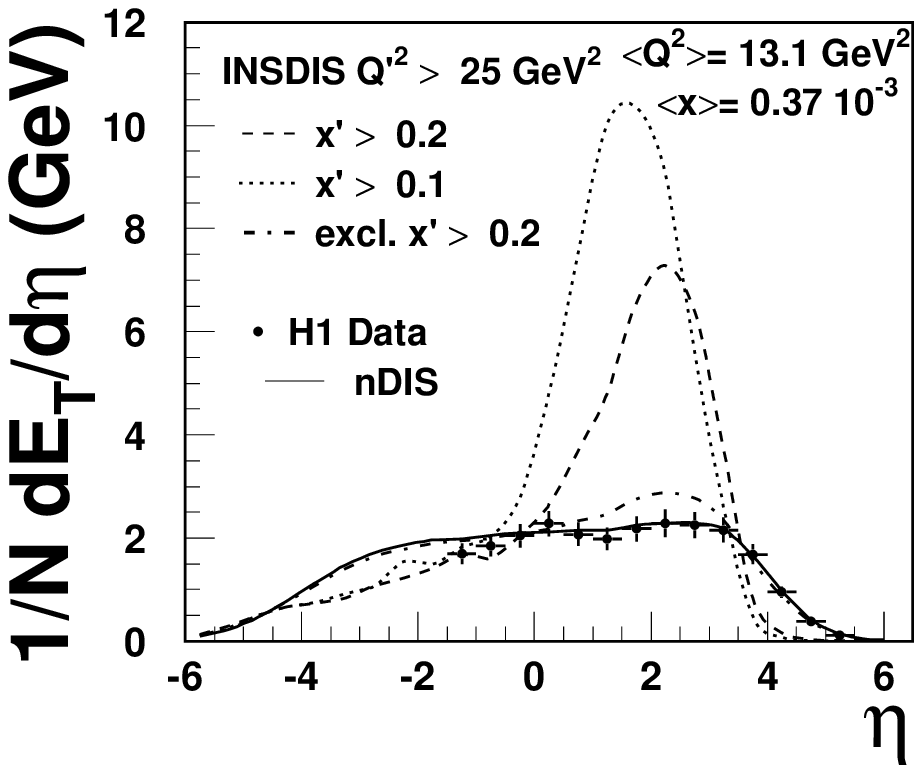,width=7cm}%
   \epsfig{file=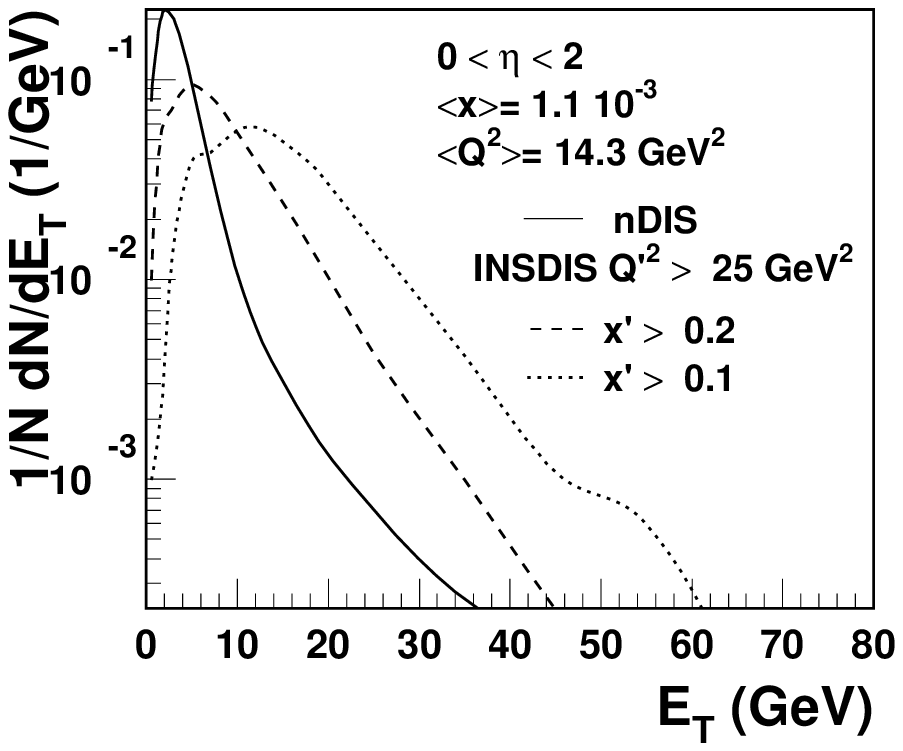,width=7cm}%
   \scaption{
             {\bf a)} The transverse energy flow \et 
             as a function of pseudorapidity $\eta$ in the hadronic CMS.
             The proton remnant direction is to the left. 
             The standard QCD model (nDIS=ARIADNE) and
             different instanton scenarios are confronted with 
             the H1 data \cite{h1:flow3}. The excluded scenario 
             with an instanton fraction $f_I> 11.8\%$ for
             $\xprime > 0.2$ is indicated.
             {\bf b)} 
             The \et distribution, where the transverse energy
             is measured in the CMS rapidity bin $0<\eta<2$,
             for two instanton scenarios, and the standard QCD model
             (nDIS).
             The plots are normalized to the total number of events $N$.}
   \label{flows} 
\end{figure}

We study DIS events in the center of mass system (CMS),
of the incoming proton and the virtual boson, i.e. the CMS of 
the hadronic final state with invariant mass $W$.
Longitudinal and transverse quantities are calculated 
with respect to the virtual boson direction (defining the $+z$ direction).

In this analysis we concentrate on the instanton production scenario
with $\xprime > 0.2$ and $\qprimesq > 25$~\GeVsq as cut-off parameters.
In this phase space region the instanton cross section calculation
is considered to be relatively safe \cite{inst:yaroslavl}. To investigate
the sensitivity to \xprime and \qprimesq we also consider scenarios
where either the event topology is 
spectacular (low \xprimex, large \qprimesqx),
or the cross section can be potentially large (low \xprimex). 

In the rest frame of the instanton induced subprocess 
partons are isotropically emitted. This multi-parton state
consists of gluons and all quark flavours
which are kinematically allowed. 
In the CMS the hadrons emerging from the instanton subprocess
occupy a homogeneous band in pseudorapidity\footnote{The pseudorapidity
$\eta$ is defined as $\eta= - \ln{\tan{\theta/2}}$, where
$\theta$ is the angle of the hadron with respect to the virtual
photon direction in the hadronic center of mass frame.}\cite{inst:vladimir}. 
As an example, Fig.~\ref{flows}a shows the flow of 
hadronic transverse energy (\etx) as a 
function of pseudorapidity $\eta$ in a certain \xb-\Qsq bin.
The height and position of the instanton
band depends on the chosen production scenario. For lower \xprime 
the instanton band gets stronger and moves towards the remnant.
In normal DIS events on average an \et of $2$~\GeV per unit pseudorapidity
is observed. In instanton induced events, the average
\et may go up to $10$~\GeV per $\eta$ unit for low \xprimex. 
A possible search strategy could involve
the \et distribution in a selected rapidity band (Fig.~\ref{flows}b),
looking for high \et events in the tail of the distribution.

\begin{figure}[h]
   \centering
   \epsfig{file=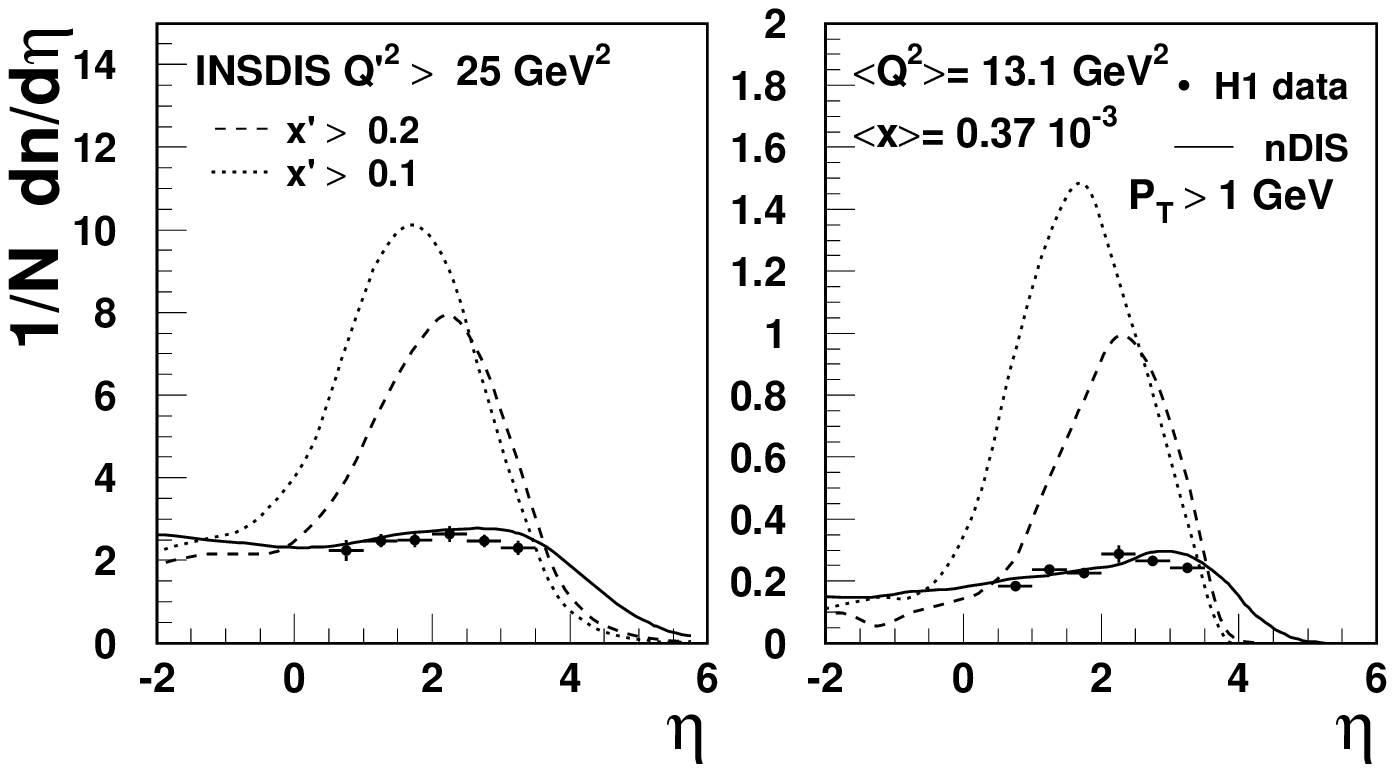,width=14cm}%
   \scaption{
             The rapidity distribution of charged particles
             in the hadronic CMS for {\bf a)} all charged particles,
             and  {\bf b)} charged particles with $\pt>1~\GeVx$.
             The proton remnant direction is to the left. 
             The standard QCD model (nDIS=ARIADNE) and
             different instanton scenarios (INSDIS) are confronted with 
             the H1 data \cite{h1:pt}. 
             The plots are normalized to the total number of events $N$.
             Both plots are for 
             $\av{\xb}= 0.37 \cdot 10^{-3}$ and
             $\av{\Qsq}= 13.1$\GeVsqx.
   }
   \label{multi} 
\end{figure}

The same effect can be seen in the pseudorapidity distribution of 
charged particles, see Fig.~\ref{multi}a. 
Instanton events are characterized by 
a large particle density localized in rapidity. In normal
DIS events there are about 2 charged particles per unit of
rapidity \cite{h1:pt}, 
while in instanton events there are up to $10$ charged
particles per unit of rapidity for a low \xprime cut-off, and somewhat less
for a high \xprime cut-off.

\begin{figure}[h]
   \centering
   \epsfig{file=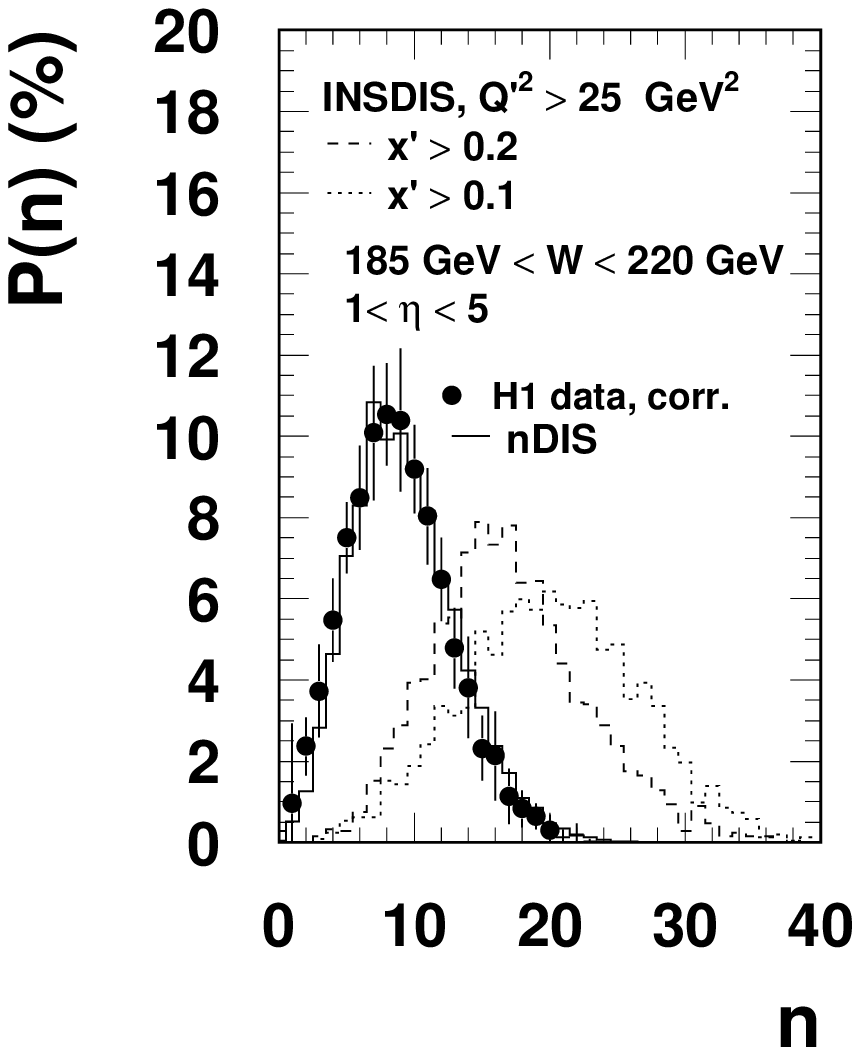,width=7cm,
    bbllx=62pt,bblly=436pt,bburx=310pt,bbury=742}
   \epsfig{file=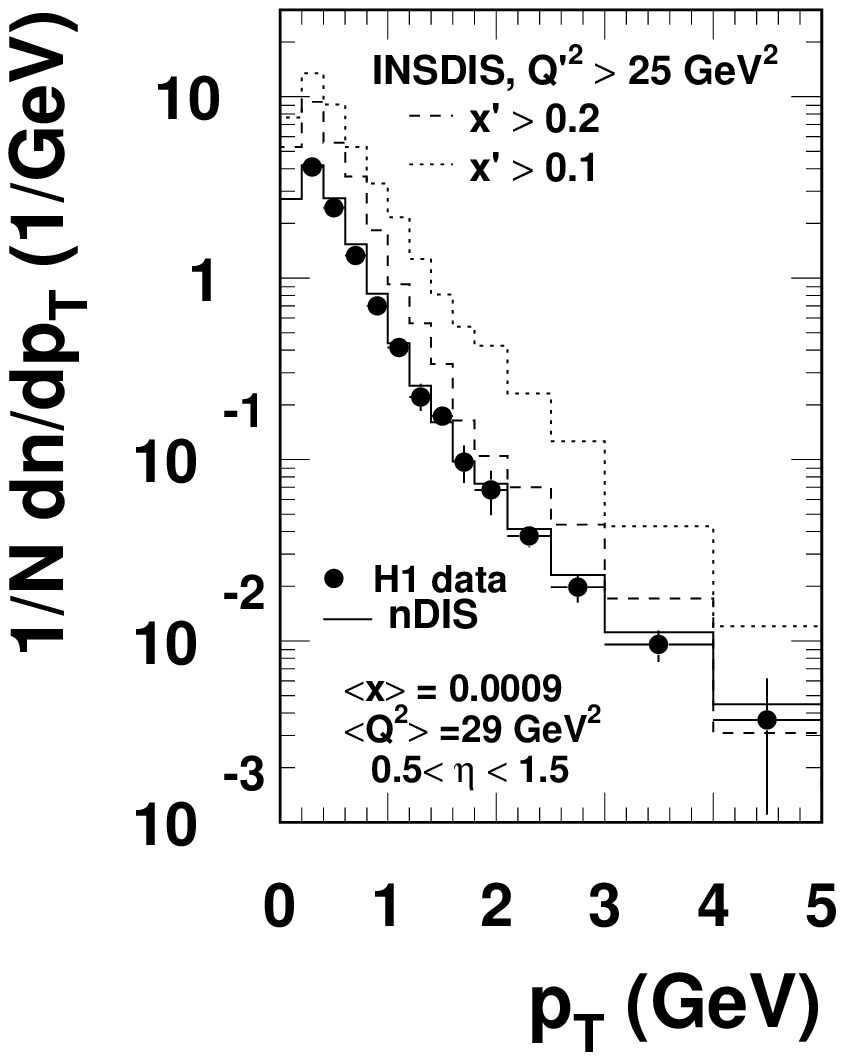,width=7cm,%
    bbllx=62pt,bblly=436pt,bburx=310pt,bbury=742,clip=}
   \scaption{ 
              {\bf a)}
              The 
              probability distribution $P(n)$ of the charged particle 
              multiplicity $n$
              from the CMS pseudorapidity range $1<\eta<5$ for
              events with $185 \GeV < W < 220~\GeVx$.
              Shown are the unfolded H1 data \cite{h1:mult},
              the expectation from a standard 
              DIS model (nDIS=Ariadne),
              and the predictions for instanton events with
              different cut-off scenarios.
              {\bf b)} 
              The transverse momentum spectrum (\ptx) of charged
              particles in the pseudorapidity range $0.5<\eta<1.5$.
              Shown are the expectations
              for normal DIS (nDIS) and for instanton events
              (INSDIS), and the 
              measurement from H1 
              \cite{h1:pt}. 
              The plot is normalized to the total number of events $N$.} 
   \label{dpdn} 
\end{figure}

The production of instanton induced events 
would be signaled by events with
abnormally large particle multiplicity.
In Fig.~\ref{dpdn} the distribution of charged particle
multiplicities is shown.
A significant fraction of the instanton events would 
lead to charged multiplicities which are very 
unlikely to be found in normal DIS events, as
predicted from standard QCD Monte Carlos.
Higher multiplicities are produced for lower \xprime and
larger \qprimesq due to the larger $M_I$.
For small \xprime also the transverse momentum (\ptx) 
spectrum of charged particles 
is somewhat harder for instanton events than for
normal DIS events (see Fig.~\ref{dpdn}b), owing to the many
semi-hard partons emerging from the instanton subprocess.
This is also reflected in the multiplicity flow of
hard particles ($\pt>1~\GeVx$), see Fig.~\ref{multi}b.
For low \xprimex, there are about $1.4$ hard charged particles
per unit rapidity in the instanton band, while for normal
DIS only 0.2 would be expected on average.

The distributions presented so far are compared to
actual HERA measurements \cite{h1:mult,h1:flow3,h1:pt}. 
The data can be described relatively well by standard
QCD models. This fact allows to place bounds
on instanton production, which will be discussed
in the next section.

\section{Limits on instanton production}

\subsection{Limits from shape comparisons}

To determine an upper limit on the instanton induced production
cross-section, it is assumed that the considered measured observable
results from an admixture of instanton induced  (INSDIS) to normal DIS (nDIS)
events with a fraction $f_I$.
From a $\chi^2$ test, comparing
$f_I \cdot {\rm INSDIS} + (1 - f_I) \cdot {\rm nDIS}$ 
events with the data, the maximally allowed fraction ($f_{lim}$)
of instanton events in the data is deduced at $95\%$
confidence level (C.L.).
The upper bound on the instanton production cross-section
($\sigma_{lim}$) is calculated from
$f_{lim}$, the number of generated instanton and 
normal DIS events and the known total DIS cross-section 
in the considered kinematic region.

Among all investigated observables in the hadronic final
state of DIS events, the transverse energy flow ($\dif E_T/\dif \eta$), 
the flow of hard charged particles ($\dif n/\dif \eta$ for $p_T > 1~\GeVx$),
and the transverse momentum spectra of charged particles in a
restricted rapidity region $0.5< \eta < 1.5$
give the best sensitivity in the analysis where shapes of data are 
compared to Monte Carlo simulations.
An example of an instanton admixture to normal DIS which can be excluded at 
$95\%$ C.L. is shown in Fig.~\ref{flows}.

\begin{figure}[htp]
\begin{center}
\begin{tabular}{cc}
 \epsfig{width=15cm,file=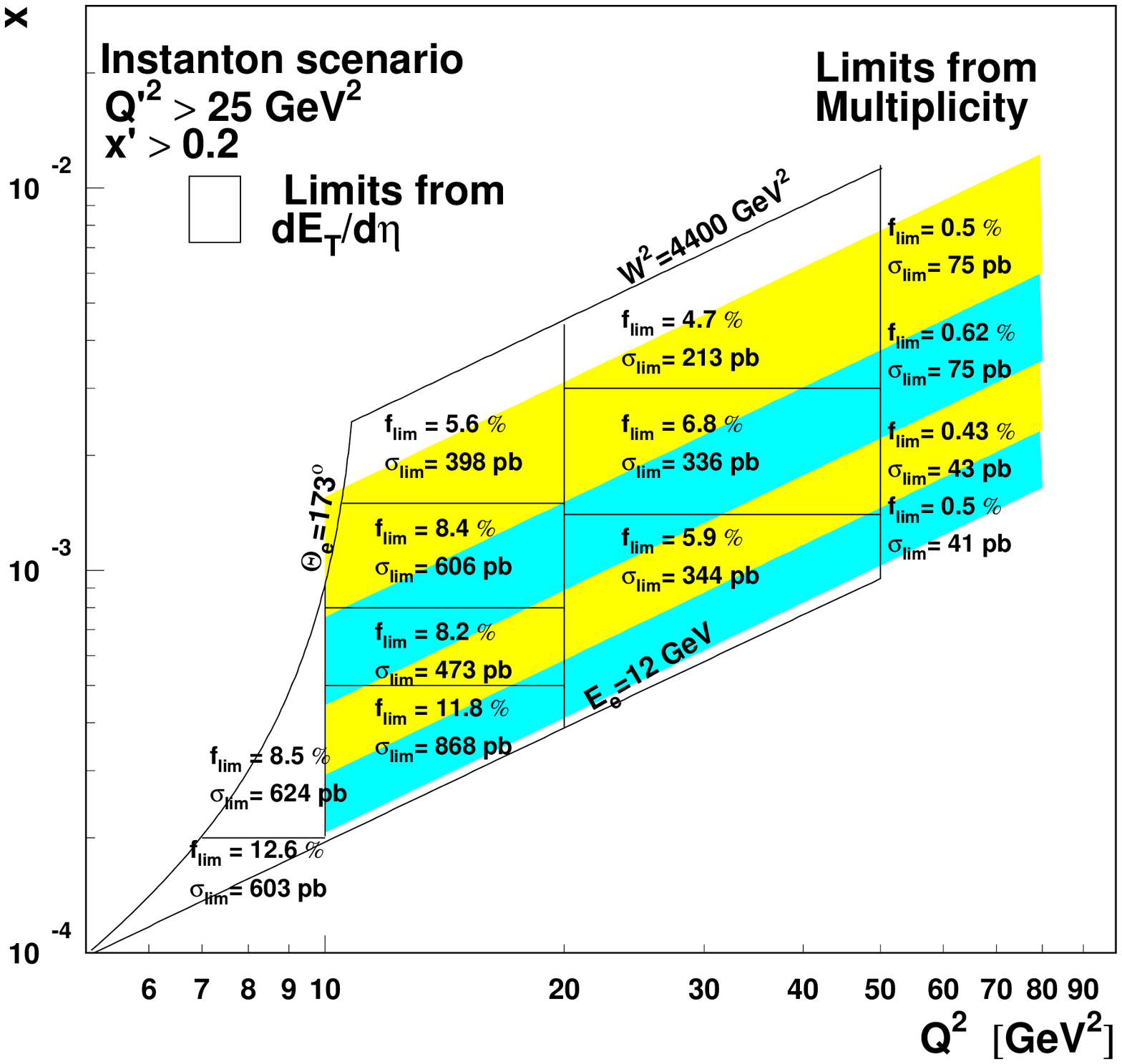,
        bbllx=28pt,bblly=142pt,bburx=561pt,bbury=668,clip=}
%
\end{tabular}
\end{center}
\vspace{-0.5cm}
 \scaption{ \label{fig:limits}
 {    Limits on instanton production with $\qprimesq > 25$\GeVsq and
      $\xprime > 0.2$.
      The cross-section limits ($\sigma_{\rm lim}$) 
      together with the maximally allowed instanton fraction
      $f_{\rm lim}$ are shown in the (\xbx,\Qsqx) plane
      obtained from the $\dif E_T/\dif \eta$ (open fields)
      and multiplicity analysis (shaded fields) with their
      numbers at the right edge.
 }}
\end{figure}

In Fig.~\ref{fig:limits} the results 
for instantons with $\xprimex > 0.2$ and $\qprimesqx > 25$\GeVsq
are summarized.
For $ 10^{-4} \lsim \xb \lsim 10^{-2}$ 
and $ 10 < \Qsq < 50$\GeVsqx,  instanton fractions $f_I$  between
$5-10$\% can be excluded from the \et flow data.
 This corresponds to cross-section limits
of $\approx 200-800$\pbx. The best limits are reached in the domain of higher
\xb and \Qsqx.
The limits obtained from the charged particle $\eta$ and \pt
spectra are of comparable order as the ones obtained from the
\et flows.
When going into the low $\xprime$ region ($\xprimex > 0.1$), where
the cross-section calculation is doubtful but the event topology
more distinct, somewhat better $f_{\rm lim}$ values of about
$3-8$\% can be reached.
Increasing the \qprimesq cut-off to $100$\GeVsq 
has a similar effect. 
However, the cross-section predicted by the instanton Monte Carlo
drops by a factor of $10$, such that the gain in sensitivity
due to the more distinct event topology is not big enough
to compensate for the falling instanton cross-section.

%

Systematic uncertainties can be investigated by
varying options in the standard DIS Monte Carlo, like the
parameterization of the proton structure function $F_2$ or
parameters associated with the hadronisation model.
Among the $F_2$ parameterizations from 
Martin, Roberts and Stirling \cite{th:mrsh} 
(MRS-H), from the CTEQ collaboration \cite{cteq4} (CTEQ-4d) and 
from Gl\"uck, Reya and Vogt \cite{th:grv} (GRV-94 HO), 
MRS-H turned out to give 
the most conservative limits and was therefore chosen to determine
$\sigma_{lim}$ and $f_{lim}$.
Using other structure function parameterizations 
leads to up to 50\% lower cross-section limits 
and, in exceptional cases, they are even lowered by a factor of $2$.
Replacing the JETSET hadronisation parameters by sets which
have been tuned to LEP data \cite{leptuning} alters
the cross-section limits by about $20$\%.

One has to keep in mind that this method relies on the assumption
that the used standard QCD Monte Carlo is a good model of 
standard QCD effects without instantons. Though it gives an
excellent description of the available hadronic
final state data over the full kinematic plane,
the true QCD dynamics, in particular at small \xbx,
is still under debate. The method discussed in the following  
section does not rely on this assumption.

%
%

\subsection{Limits from multiplicity distributions}

The most sensitive data to instanton induced processes are the 
multiplicity distributions $P(n)$ giving the probability
to produce $n$ hadrons in an event.
Instanton events are characterized by abnormally large
hadron multiplicities $n$.
H1 has presented the observed
charged particle multiplicity distribution in
the CMS pseudo-rapidity interval $1<\eta<5$ for four
different regions of $W$ \cite{h1:mult}. 
Here we exploit the fact that
events above a certain multiplicity
\nmax were not observed to
place limits on instanton production 
\footnote{H1 already derived a limit on instanton production
by looking for a deviation from the multiplicity distribution
for normal DIS events, assumed to be a negative binomial distribution,
for multiplicities $n<\nmax$\cite{h1:mult}.}. 

The observation of no events above \nmax has to be compared
to the expectation from instanton events. We define the 
instanton search region in the multiplicity distribution
by requiring $n>\nmax$. 
The fraction of instanton events satisfying the multiplicity
cut $n>\nmax$ gives the instanton efficiency \einst 
\begin{equation}
\einst(\nmax) = \sum_{\mbox{$n$} > \nmax} P_I(n)
\label{eq:einst}
\end{equation}  
and
is calculated with the instanton generator.

By comparing the observed and the unfolded data multiplicity
distribution from H1 we conclude that they are very similar;
if anything, the raw distribution is 
slightly broader than the unfolded
distribution.
Therefore the true instanton efficiency 
including detector effects, is
very similar, and if anything, slightly larger than our
estimate from the instanton generator, which neglected
detector effects. Our assumption on \einst~ is therefore
conservative. We assume however, that the H1 acceptance is not
biased against high multiplicity events.
This appears reasonable as H1 imposes only very loose cuts
in the DIS event selection \cite{h1:mult}. For example, the requirement
of a reconstructed event vertex from charged tracks should
introduce losses only for low multiplicity events.
Also, tracks from instanton events
are produced isotropically,
avoiding very dense track configurations with potential
problems due to the limited double track resolutions in
the central drift chamber.

The same line of reasoning applies to QED radiative effects,
because for the unfolded distributions also radiative corrections
had been applied. Furthermore, with the given event selection
and kinematic reconstruction method, the kinematic distributions
are little affected by QED radiation \cite{h1:hess}.

The 95\% C.L. limit \flim~on the fraction of instanton events
in a given DIS event sample $f_I$ is then calculated from the 
95\% C.L. upper limit corresponding to zero events seen
(i.e. three), \einst, and the total number of DIS events observed $N$
(given by H1):
\begin{equation}
f_I < \flim = \frac{3/\einst}{N} \hspace{0.5cm} (95\% \; {\rm C.L.}) 
\label{eq:flim}
\end{equation}  
This fractional limit can be converted into a cross-section limit:
\begin{equation}
\siginst < \siglim = \flim \cdot \sigdis 
\label{eq:siglim}
\end{equation}  
Here \sigdis is the DIS cross-section calculated for the 
kinematic bin from a recent parameterization of the 
parton density functions \cite{th:mrsh} which fits the
HERA structure function data \cite{h1:f2of94,z:f2}.
The H1 data used are summarized in table ~\ref{tab:multi_h1},
and the results from 
this analysis are presented in table~\ref{tab:lim_multi}
and Fig~\ref{fig:limits}.
Contrary to previous analyses \cite{h1:mult,h1:k0} 
and the other results discussed in this paper, these results do not 
depend on models for the standard DIS process, because no
background had to be subtracted.

\begin{footnotesize}
\begin{table}
\begin{center} 
\begin{tabular}{|c|r|c|c|r|c|}
  \hline
  kinematic bin & $W (\GeVx)$ & $\av{x}/10^{-3}$ & 
  \av{\Qsqx} (\GeVsqx) & $N$ & \nmax  \\
  \hline
  A &  80-115 & 2.43 & 22.9 & 16680 & 20  \\
  B & 115-150 & 1.33 & 23.2 & 14983 & 24  \\
  C & 150-185 & 0.84 & 23.3 & 12191 & 25  \\
  D & 185-220 & 0.58 & 23.5 &  9255 & 26  \\
  \hline 
\end{tabular}
\end{center} 
\scaption{Summary of H1 multiplicity data \cite{h1:mult} which are
           used in this analysis. The data correspond to an integrated
           luminosity of 1.3 \pbinvx.}
\label{tab:multi_h1}   
\end{table} 
\begin{table}
\begin{center} 
\begin{tabular}{|c|c|c|c|c|c|}
  \hline
  \multicolumn{6}{|c|}{$\qprime^2>25 \GeVsqx$, $\xprime>0.1$} \\
  \hline
  kinematic bin & \sigdis (pb) & 
  \einst & \flim & \siglim (pb) & extrapolated \siginst (pb) \\
  \hline
  A & 14800 & 0.08   & 0.0023 & 34 &  900 \\
  B & 12200 & 0.07   & 0.0029 & 35 & 1400 \\   
  C &  9900 & 0.13   & 0.0020 & 20 & 1400 \\   
  D &  8300 & 0.19   & 0.0017 & 15 & 1300 \\   
  \hline 
  \hline
  \multicolumn{6}{|c|}{$\qprime^2>25 \GeVsqx$, $\xprime>0.2$}\\
  \hline
  kinematic bin & \sigdis (pb) & 
  \einst & \flim & \siglim (pb) & predicted \siginst (pb) \\
  \hline
  A & 14800 & 0.04   & 0.0050 & 75 & 2.6 \\
  B & 12200 & 0.03   & 0.0062 & 75 & 3.1 \\   
  C &  9900 & 0.06   & 0.0043 & 43 & 2.9 \\   
  D &  8300 & 0.07   & 0.0050 & 41 & 2.7 \\   
  \hline 
\end{tabular}
\end{center} 
\scaption{Limits on instanton production from the analysis of
           the H1 multiplicity data compared to predicted
           and extrapolated instanton production cross section.
           The prediction is only considered reliable for
           $\xprime \gsim 0.2$.
         }
\label{tab:lim_multi}   
\end{table} 
\end{footnotesize}

The predicted instanton induced cross-sections crucially depend on 
the \xprime and \qprime cut-offs,
and on the renormalisation and factorization scheme  chosen at this
level of approximation, varying by 
orders of magnitudes for the chosen scenarios.
\einst~and the derived instanton limits depend much less 
upon the instanton kinematics, they typically vary by a factor $2$ or $3$.
These findings are summarized in table~\ref{tab:scenarios}.
Experimentally favorable are large \qprimesq and low \xprimex,
because that results in a large instanton ``mass''.
The instanton induced cross-section decreases with increasing
\qprimesq and \xprime. 
For lower \xprime the cross-section calculation becomes unreliable,
because higher order interactions are expected
to dampen the growth in cross-section.
The predictions quoted here are calculated with the
instanton Monte Carlo, which extrapolates also into the
unreliable region of small \xprime and \qprimesqx.

\begin{figure}[htb]
\begin{center}
\begin{tabular}{cc}
 \mbox{
 \epsfig{width=7.5cm,file=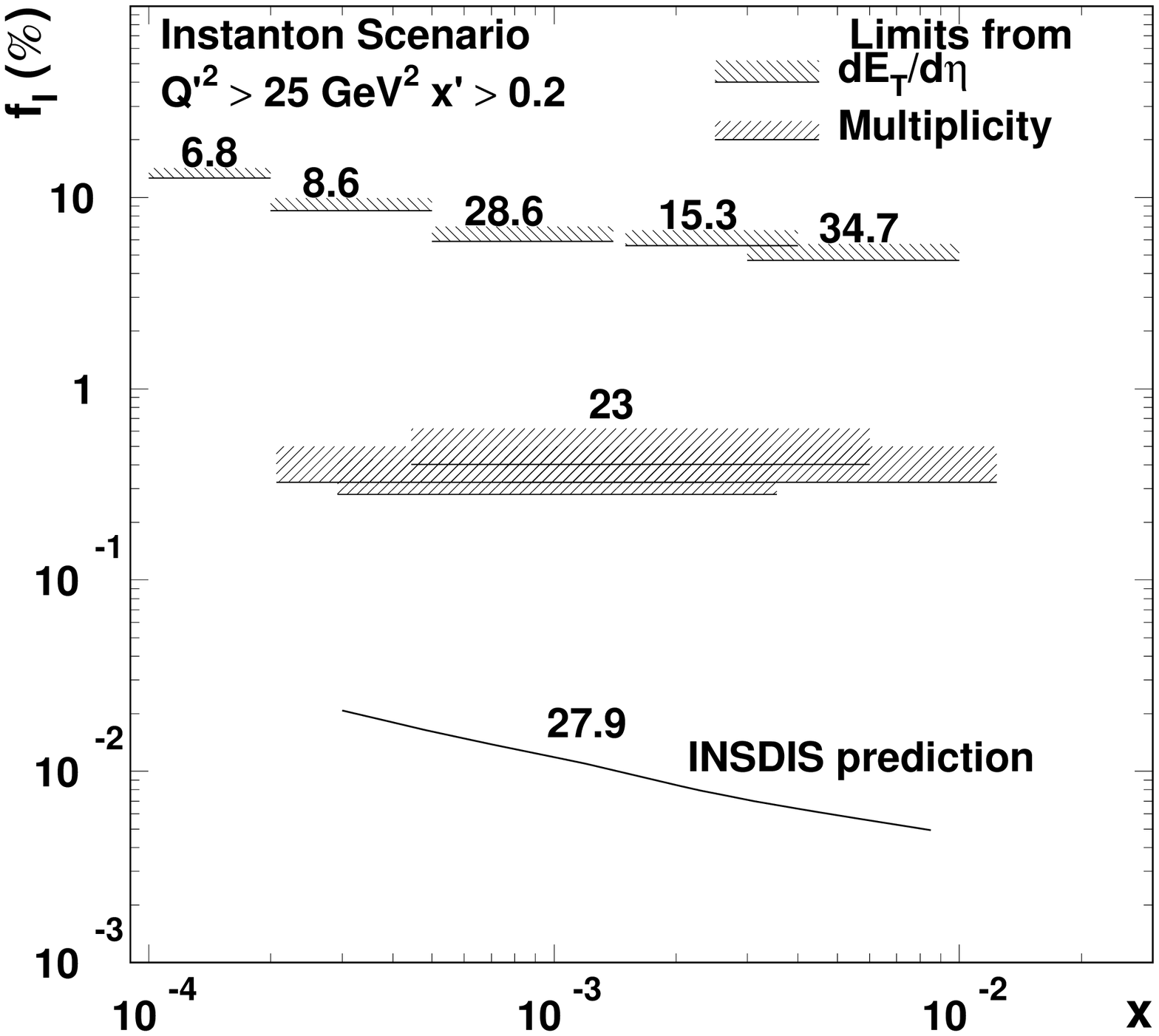}
 }
 \mbox{
  \epsfig{width=7.5cm,file=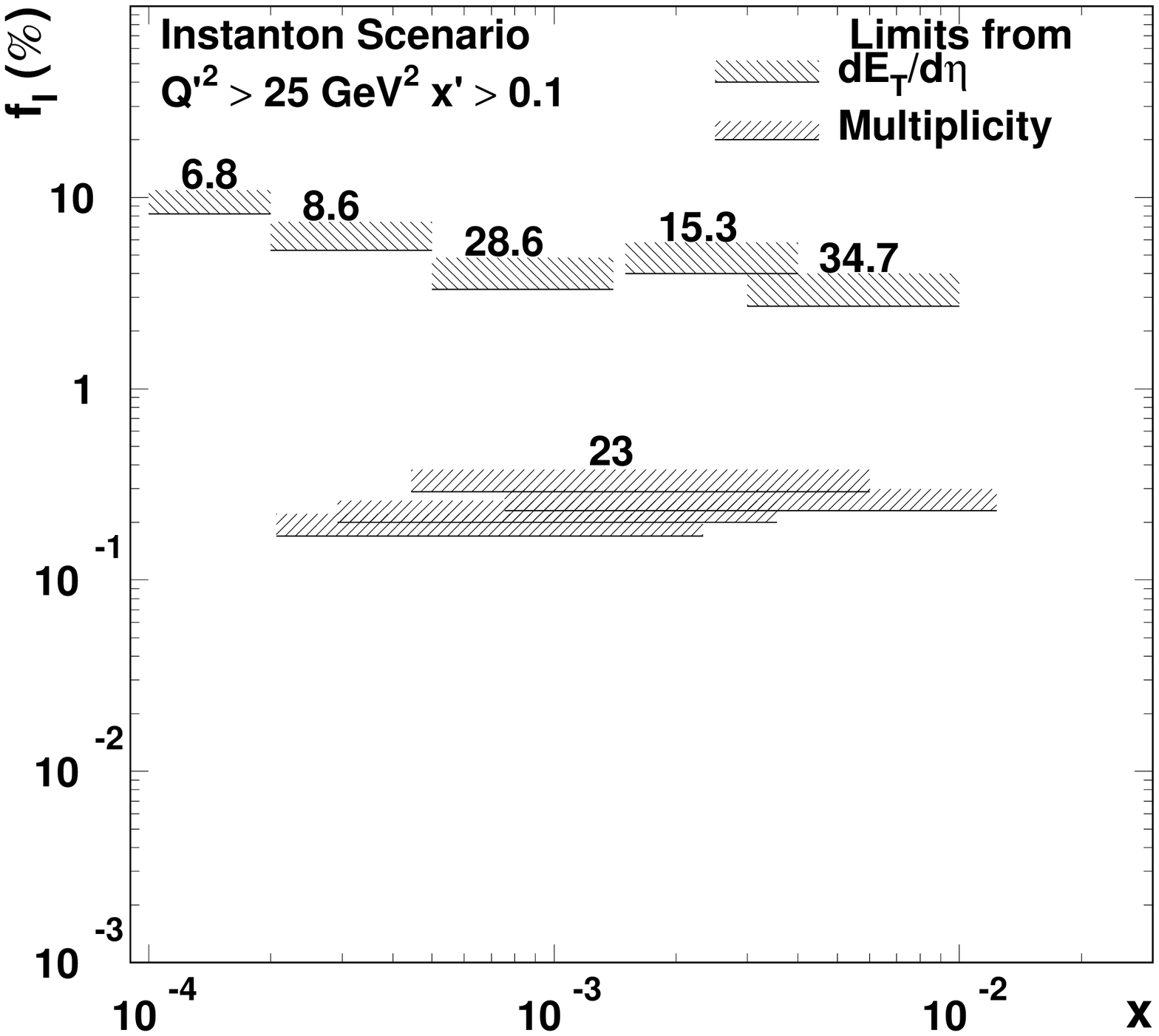}
 }
%
\end{tabular}
\end{center}
\vspace{-0.5cm}
 \scaption{ \label{fig:limfra}
 {    The maximally allowed fraction $f_{\rm lim}$
      of instanton induced events in DIS for 
      a) $\qprimesq > 25$\GeVsq and $\xprime > 0.2$ and 
      b) $\qprimesq > 25$\GeVsq and $\xprime > 0.1$  
      from transverse energy flows and the multiplicity distribution 
      as function of $x$. Regions above the lines are excluded
      at 95\% C.L..
      The numbers give the average \Qsq values in \GeVsq for
      the $x$ bins. 
      The theory prediction for $10 \GeVsq < \Qsq < 80$~\GeVsq
      is superimposed (full line).
 }}
\end{figure}

For a theoretically ``safe'' scenario, 
$\qprimesq>25 \GeVsq$ and $\xprime>0.2$ \cite{inst:yaroslavl},
the limits are still roughly a factor $20$ away from the 
predicted cross-section (see Fig.~\ref{fig:limfra}).
When the \xprime cut-off is lowered to $0.1$, the instanton
efficiency improves by roughly a factor two, and the
extrapolated instanton cross-section increases by a factor
$\approx 500$. 
The data clearly rule out such a large
instanton cross-section and thus provide a constraint
for the behavior of the functions $F(\xprimex)$ and $\Sigma(\xprimex)$
(see equation~\ref{eq:cross})
in the theoretically uncertain region at small $\xprimex$.
Qualitatively the same features are observed for
the other \qprimesq scenarios
with a very low and a very high cut-off.

\begin{footnotesize}
\begin{table}
\begin{center} 
\begin{tabular}{|ll|rr|rr|rr|rr|}
  \hline
     \multicolumn{2}{|c|}{Instanton cross-section (pb)} & 
     \multicolumn{2}{c|}{bin A} & 
     \multicolumn{2}{c|}{bin B} & 
     \multicolumn{2}{c|}{bin C} & 
     \multicolumn{2}{c|}{bin D}  \\
    \multicolumn{2}{|c|}{limit/predicted} &
\siglim & \siginst & \siglim & \siginst & \siglim & 
 \siginst & \siglim & \siginst \\
  \hline
$\qprimesq>0.05\GeVsq$&$\xprime>0.1$ & 43&1500 & 49&2200 & 22&2300 & 17&2000\\
$\qprimesq>0.05\GeVsq$&$\xprime>0.2$ & 97&7.1 & 124&7.3 &  66&6.9 &  51&5.7 \\
\hline
$\qprimesq>25\GeVsq$&$\xprime>0.1$ & 34&930 & 35&1400 & 20&1400 & 15&1300 \\
$\qprimesq>25\GeVsq$&$\xprime>0.2$ & 75&2.6 & 75&3.1  & 43&2.9  & 41&2.7 \\
\hline
$\qprimesq>100\GeVsq$&$\xprime>0.1$ & 31&140 & 23&270 & 12&300 & 10&330 \\
$\qprimesq>100\GeVsq$&$\xprime>0.2$ & 28&0.3 & 23&0.5 & 13&0.5 & 13&0.4 \\
\hline
\end{tabular}
\end{center} 
\scaption{
           Cross-section limit \siglim and predicted/extrapolated
           cross-section \siginst for the different (\xprimex,\qprimesqx)
           scenarios.
           The prediction is only considered reliable for
           $\xprime \gsim 0.2$.
          }
\label{tab:scenarios}   
\end{table} 
\end{footnotesize} 

\section{Conclusions}

The observation of instanton effects in DIS events at HERA
would be a novel, non-perturbative manifestation of QCD
and would furthermore provide valuable indirect information about
$B+L$ violation in the multi-TeV region induced by electroweak
instantons.
The distinct event topology of instanton induced events allows
to discriminate them from normal DIS events.
Using existing HERA data on the hadronic final state 
corresponding to an integrated luminosity of ${\cal O}(1\pbinvx)$,
the maximally allowed fraction of instantons in DIS is found to 
be of ${\cal O}(1\%)$ for $ 80 < W < 220$\GeV
and $\xprime>0.2$ and $\qprimesq > 25$\GeVsqx.
In this phase space region the predicted instanton fraction
is $\approx 0.01-0.02\%$, i.e. still below
the level excluded by existing HERA data.
Dedicated instanton searches employing more elaborate search strategies
and higher luminosities will help to test the 
prediction of the cross-section in future.

%
%

\section{Acknowledgements}

We would like to thank A. Ringwald and F. Schrempp for many
inspiring discussions on the instanton theory,
P. van Mechelen for discussions on multiplicity distributions,
and our summer students B. Koblitz
and C. Tesch for their help. 
We also thank J. Gayler, A. Ringwald and F. Schrempp for their critical
reading of the manuscript.
M. K. wishes to thank the
Deutsche Forschungsgemeinschaft for their support.

\begin{footnotesize}
%
%

\end{footnotesize}
\end{document}